\documentclass[superscriptaddress,floatfix,prl,twocolumn,amsmath,amssymb]{revtex4-1}

\usepackage{graphicx}
\usepackage{dcolumn}
\usepackage{bm}
\usepackage{amssymb}
\usepackage{natbib}
\usepackage{amsmath}
\usepackage{mathtools}
\usepackage{color}
\usepackage{datetime}
\usepackage{footnote}
\usepackage{bbold}
\usepackage[normalem]{ulem}
\usepackage{dcolumn}
\usepackage{ragged2e}
\usepackage{xfrac}
\usepackage{sidecap}
\usepackage{setspace}
\usepackage{multirow}
\usepackage{xcolor}
\usepackage{tikz}
\usepackage{tikz-cd}
\usepackage{booktabs}
\usepackage{flushend}

\usepackage[ colorlinks = true,
             linkcolor = blue,
             urlcolor  = blue,
             citecolor = red,
             anchorcolor = green,
]{hyperref}

\usepackage{qcircuit}

\newcommand{\ket}[1]{\ensuremath{|#1\rangle}}
\newcommand{\bra}[1]{\ensuremath{\langle#1|}}

\newcommand{\be}{\begin{equation}}
\newcommand{\ee}{\end{equation}}

\newcommand{\nocontentsline}[3]{}
\newcommand{\tocless}[2]{\bgroup\let\addcontentsline=\nocontentsline#1{#2}\egroup}

\newcommand{\tr}{\mathrm{tr}}
\newcommand{\eps}{\varepsilon}
\newcommand{\cE}{\mathcal{E}}

\newif\iffigs
\figstrue

\begin{document}

\title{Beating the Classical Limits of Information Transmission using a Quantum Decoder}

\author{Robert~J.~Chapman}
\affiliation{Quantum Photonics Laboratory and Centre for Quantum Computation and Communication Technology, School of Engineering, RMIT University, Melbourne, Victoria 3000, Australia}
\affiliation{School of Physics, The University of Sydney, Sydney, New South Wales 2006, Australia}

\author{Akib Karim}
\affiliation{Quantum Photonics Laboratory and Centre for Quantum Computation and Communication Technology, School of Engineering, RMIT University, Melbourne, Victoria 3000, Australia}
\affiliation{School of Physics, The University of Sydney, Sydney, New South Wales 2006, Australia}

\author{Zixin Huang}
\affiliation{Quantum Photonics Laboratory and Centre for Quantum Computation and Communication Technology, School of Engineering, RMIT University, Melbourne, Victoria 3000, Australia}
\affiliation{School of Physics, The University of Sydney, Sydney, New South Wales 2006, Australia}

\author{Steven~T.~Flammia}
\affiliation{Centre for Engineered Quantum Systems, School of Physics, The University of Sydney, Sydney, New South Wales 2006, Australia}
\affiliation{Center for Theoretical Physics, Massachusetts Institute of Technology, Cambridge, USA}

\author{Marco Tomamichel}
\affiliation{Centre for Engineered Quantum Systems, School of Physics, The University of Sydney, Sydney, New South Wales 2006, Australia}
\affiliation{Centre for Quantum Software and Information, School of Software, University of Technology Sydney, Sydney, New South Wales 2007, Australia}

\author{Alberto Peruzzo}
\email{alberto.peruzzo@rmit.edu.au}
\affiliation{Quantum Photonics Laboratory and Centre for Quantum Computation and Communication Technology, School of Engineering, RMIT University, Melbourne, Victoria 3000, Australia}
\affiliation{School of Physics, The University of Sydney, Sydney, New South Wales 2006, Australia}

\begin{abstract}
Encoding schemes and error-correcting codes are widely used in information technology to improve the reliability of data transmission over real-world communication channels. 
Quantum information protocols can further enhance the performance in data transmission by encoding a message in quantum states; however, most proposals to date have focused on the regime of a large number of uses of the noisy channel, which is unfeasible with current quantum technology. 
We experimentally demonstrate quantum enhanced communication over an amplitude damping noisy channel with only two uses of the channel per bit and a single entangling gate at the decoder. 
By simulating the channel using a photonic interferometric setup, we experimentally increase the reliability of transmitting a data bit by greater than 20\% for a certain damping range over classically sending the message twice. 
We show how our methodology can be extended to larger systems by simulating the transmission of a single bit with up to eight uses of the channel and a two-bit message with three uses of the channel, predicting a quantum enhancement in all cases.
\end{abstract}

\maketitle

\section{Introduction}

Data transmission is an indispensable resource in information technology and requires reliable communication over realistic, noisy channels.
Information can be protected against noise by adding redundancy---for example, sending multiple copies of each bit---at the cost of reducing the data transmission rate (in transmitted bits per use of the channel).
Encoding each bit in an optimal basis can increase the transmission rate up to the \textit{channel capacity}, where the information can be decoded with negligible error; however, this usually requires large numbers of uses of the channel \cite{shannon_mathematical_1948}.
To increase the transmission rate beyond the channel capacity, we can encode information in quantum states and perform coherent joint measurement across all the qubits \cite{guha_structured_2011, buck_experimental_2000, sasaki_quantum_1998, tomamichel_second-order_2015, chubb_2017, cheng_2017, wilde_second-order_2015, chen_optical_2012, nair_2014, lloyd_sequential_2011, rosati_achieving_2016, wang_one-shot_2012, brandao_entangled_2011} to reach the \textit{Holevo capacity} \cite{holevo_bounds_1973, schumacher_sending_1997, holevo_capacity_1998}.

The quantum capacity of a noisy channel is only applicable in the regime of asymptotically many uses of the channel, which requires coherent control of asymptotically many qubits \cite{tomamichel_quantum_2016}.
This is unrealistic for current quantum technology and, therefore, a different approach is necessary to find quantum enhanced robustness to noisy channels with limited resources.
In this setting, we can no longer seek error-free communication, but work to minimize the probability of inevitable errors.
While the advantages of quantum states to increase the channel capacity has been reported for the amplitude damping channel \cite{giovannetti_information-capacity_2005, darrigo_classical_2013, darrigo_information_2015, jahangir_quantum_2015} and other noisy channels, it is less well known whether such a quantum enhancement exists when operating far from the asymptotic regime. 
General bounds in the one-shot regime are given in Refs.~\cite{wang_one-shot_2012, brandao_entangled_2011} but are loose (upper and lower bounds differ by several bits) when only a few uses of the amplitude damping channel are considered.
Moreover, to date most investigations in this direction consider quantum schemes where a portion of the system is immune to the noisy channel, for example, a noiseless shared entangled state~\cite{prevedel_entanglement-enhanced_2011, hemenway_optimal_2013, williams_entanglement_2011} or a noiseless ancilla qubit to assist a noisy one \cite{ghalaii_quantum_2016}. 
Quantum enhancement has been numerically demonstrated for transmitting a one-bit message with two uses of a Pauli channel \cite{bennett_entanglement-enhanced_1997}.

\begin{figure*}
\centering
\includegraphics[width=1.0\linewidth]{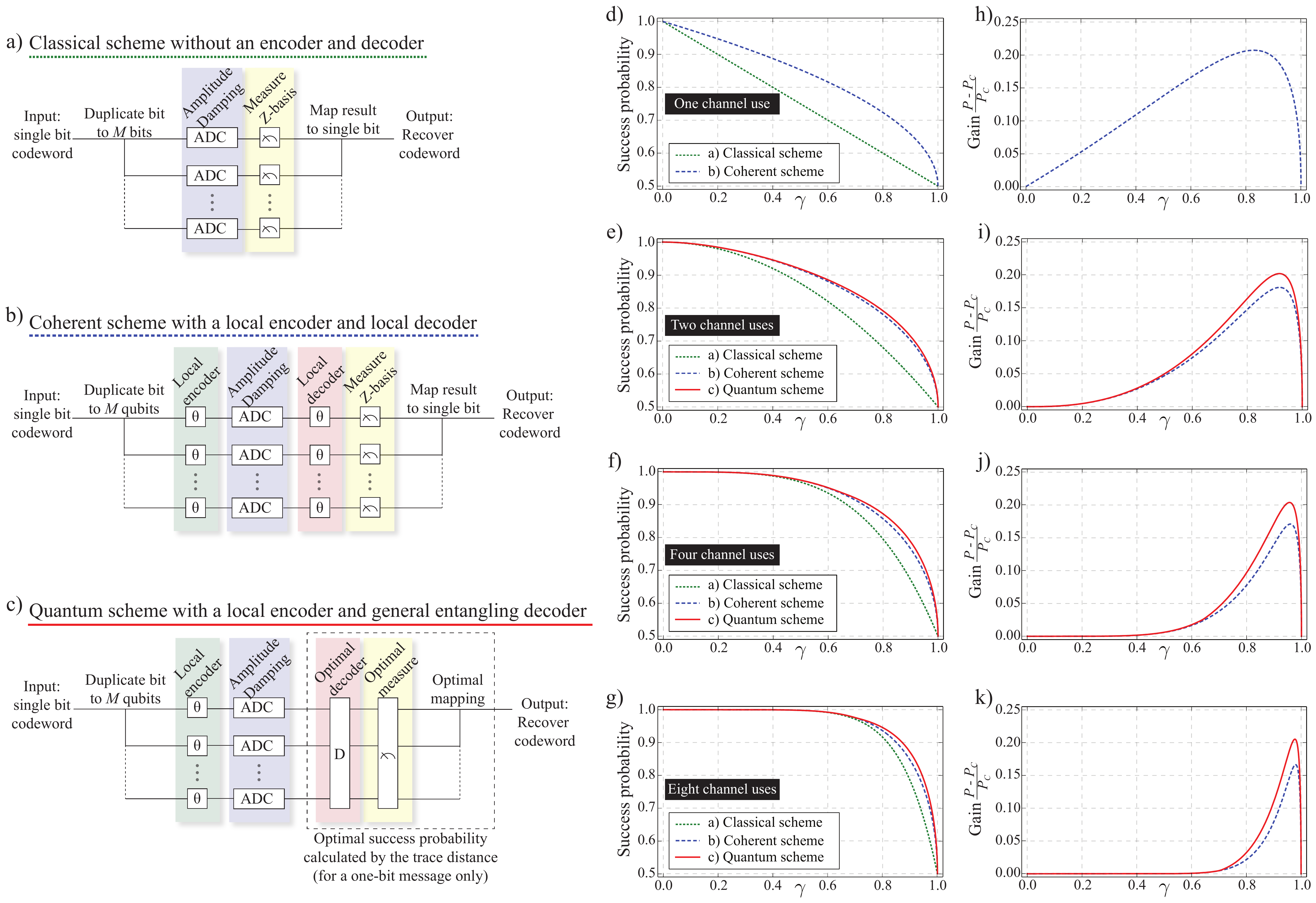}
\caption{
Numerical simulations for transmitting one bit over an amplitude damping channel encoded in the state of one, two, four and eight qubits.
(a) The classical scheme of duplicating a bit $M$ times before transmission over an ADC. 
(b) The coherent scheme that uses separable coherent quantum states to encode the data with a local rotation $\theta_\gamma$, which is numerically optimized for each $\gamma$ value. 
(c) The quantum scheme with local encoding and an entangling measurement at the decoder. For this one-bit (two codeword) case, the trace distance between code words after the ADC yields the maximum success probability of an optimal quantum decoder, measurement, and mapping.
(d)-(g) The success probability for the classical, coherent and quantum schemes for $\gamma=[0,1]$ and $M = 1,2,4$, and $8$. 
(h)-(k) The corresponding gain over the classical scheme, calculated as $\tfrac{P-P_c}{P_c}$, where $P$ is the quantum or coherence success probability and $P_c$ is the classical probability.
}
\label{fig:fig_1}
\end{figure*}

Here, we propose and experimentally demonstrate a scheme for quantum-enhanced transmission over an amplitude damping channel, where each bit is transmitted as duplicate qubits that are entangled at the decoder after the channel.
We implement the amplitude damping channel and entangling decoder using polarization photonic qubits and we experimentally demonstrate a greater than $20\%$ enhancement in the success probability of message recovery compared to a corresponding classical scheme.
We numerically investigate encoding each bit in up to eight qubits and demonstrate that a fully quantum, entangling decoder in all cases improves the message recovery over the same number of classical or coherent but separable channel uses.
Finally, we extend our methodology for transmitting a two-bit message with three uses of the noisy channel and demonstrate that an entangling decoder enables us to beat the best classical strategy (with optimal mapping between the message and physically transmitted codeword) by more than $50\%$ and the optimal coherent scheme by more than $10\%$ for a certain damping parameter range.
Our results offer a practical approach to quantum-enhanced data transmission in the regime of minimal resources, showing an improvement over equivalent classical resources.

\section{Theoretical success probability for transmitting a one-bit message over an amplitude damping channel}

Amplitude damping is the process of asymmetric relaxation in a quantum system, such as spontaneous emission observed in trapped ions \cite{blinov_quantum_2004} and quantum dots \cite{gerardot_optical_2008}, and is a key noise process in quantum information \cite{nielsen_quantum_2011}.
The single-qubit amplitude damping channel (ADC) is given as
\begin{align}
    \mathcal{E}_{\mathrm{ad}}^{\gamma}(\rho) = \sum_i E_i\rho E_i^{\dagger}, \label{eq:ADCStates}
\end{align}
where the Kraus operators $E_i$ for the channel are 
\begin{align}
    E_0 =   \begin{bmatrix}
            1 & 0 \\
            0 & \sqrt{1-\gamma}
            \end{bmatrix}, \,\,
    E_1 =   \begin{bmatrix}
            0 & \sqrt{\gamma} \\
            0 & 0
            \end{bmatrix}.\nonumber
\end{align}
The channel $\mathcal{E}_{\mathrm{ad}}^{\gamma}$ incoherently damps the state $\ket{1}$ to $\ket{0}$ with probability $\gamma$ (the damping parameter), but leaves the state $\ket{0}$ unaffected. 
Relaxation in superconducting circuits is observed as amplitude damping \cite{friis_coherent_2015} and can limit the usable lifetime of the qubits \cite{takita_experimental_2017}.
An ADC can also describe finite squeezing in measurement-based quantum computing \cite{alexander_measurement-based_2017} and infidelity in the perfect state transfer protocol \cite{bose_quantum_2003}.

A classical (incoherent) bit $\{(0),(1)\}$ with a uniform prior has a $1 - \sfrac{\gamma}{2}$ average probability of being read correctly after transmission over an ADC.
Transmitting $M$ copies of each bit increases the probability of success at the cost of reduced information transmission rate.
The maximum classical success probability for a single bit is $P_c = 1 - \sfrac{\gamma^M}{2}$, which requires a final measurement mapping where if any of the $M$ bits are measured as $(1)$, then the original data bit is known to be $(1)$.
This is the best mapping as the channel is asymmetric in the computational basis and therefore this is the optimal \textit{classical scheme}.
We have shown this scheme in Fig.~\ref{fig:fig_1}(a) and the probability of success for $M = 1, 2, 4$ and $8$ are plotted in green in Figs.~\ref{fig:fig_1}(d)-(g).

We next consider encoding in $M$ duplicate separable superposition states, such as the polarization of laser light.
Local operations can be applied such that the encoded states are in the most robust basis for the particular noise channel, $\{\ket{\psi_0}^{\otimes M},\ket{\psi_1}^{\otimes M}\}$.
We describe this encoding as a \textit{coherent scheme} with each use of the channel comprising of a qubit; however, entanglement is not used.
We use numerical optimization to find the optimal encoding rotation $\theta_\gamma$ for each $\gamma$ value.
This rotation is applied before the channel to encode the message and after the channel to decode the data before a Z-basis projection measurement and finally a mapping, where if any of the qubits are measured as $\ket{1}$, then it is known that the original bit message was $(1)$.
This scheme is shown in Fig.~\ref{fig:fig_1}(b) and the success probabilities plotted in blue in Figs.~\ref{fig:fig_1}(d)-(g).
The optimal coherent scheme increases the success probability over classical schemes for all $\gamma$ values and all $M$, achieving a maximum gain of 20.71\% with a single channel use ($M = 1$) at $\gamma=0.830$, where the optimal encoding rotation is a Hadamard gate.
The gain is plotted in Figs.~\ref{fig:fig_1}(h)-(k) and is calculated as $\tfrac{P_{\mathrm{cohere}}-P_c}{P_c}$, where $P_{\mathrm{cohere}}$ is the success probability of the coherent scheme and $P_c$ is the classical success probability. 
As the number of channel uses increases, the maximum advantage of the coherent scheme decreases.
Also, the maximum gain is achieved at higher $\gamma$ values for more uses of the channel.

Finally, we consider all possible decoders, including entangling decoders, to maximize the success probability.
It is key that the decoder can discriminate between the basis states after the ADC, $\{\rho_0^{\otimes M},\rho_1^{\otimes M}\}$, which are no longer orthogonal.
The problem of differentiating quantum states has been addressed theoretically \cite{Helstrom1967} and experimentally \cite{higgins_mixed_2009} for a depolarizing channel using only local measurements, i.e., a coherent scheme.
The distinguishability of two quantum states can be calculated as the trace distance \cite{nielsen_quantum_2011, gilchrist_distance_2005, bennett_entanglement-enhanced_1997}
\begin{equation}
    D(\rho_0,\rho_1) = \frac{\mathrm{Tr}|\rho_0-\rho_1|}{2},
    \label{eq:D}
\end{equation}

\noindent from which the probability of successfully decoding the encoded bit is calculated as 
\begin{equation}    
    P_{\mathrm{quant}} = \frac{1 + D(\rho_0,\rho_1)}{2}.
    \label{eq:Pr}
\end{equation}

\noindent To achieve this success probability requires the optimal entangling measurement, which will differ for all $\gamma$ values.
We use a numerically optimized local encoder and the trace distance to find the optimal \textit{quantum scheme} as shown in Fig.~\ref{fig:fig_1}(c) with success probabilities plotted in red in Figs.~\ref{fig:fig_1}(e)-(g).
The quantum decoder enables an even higher success probability than the classical and coherent schemes for all $\gamma$ values and all $M$.
The gain over the classical scheme is calculated as $\tfrac{P_{\mathrm{quant}}-P_c}{P_c}$ and is plotted in Figs.~\ref{fig:fig_1}(i)-(k).
The maximum gain increases with $M$ and, for $M=8$, we calculate a gain of 20.53\% over the classical scheme at $\gamma=0.977$.

We have shown that using an entangling decoder after an ADC can enhance successful message recovery over the optimal classical and coherent schemes, however, in this numerical study we have used the trace distance to calculate the success probability which may require projective measurements that are greater than rank-one and thus impractical experimentally.
In order to experimentally achieve, or approximate, the optimal success probability, we must design a suitable entangling decoder and projection measurement that can be implemented in the laboratory.

\begin{figure*}
\centering
\includegraphics[width=1.0\linewidth]{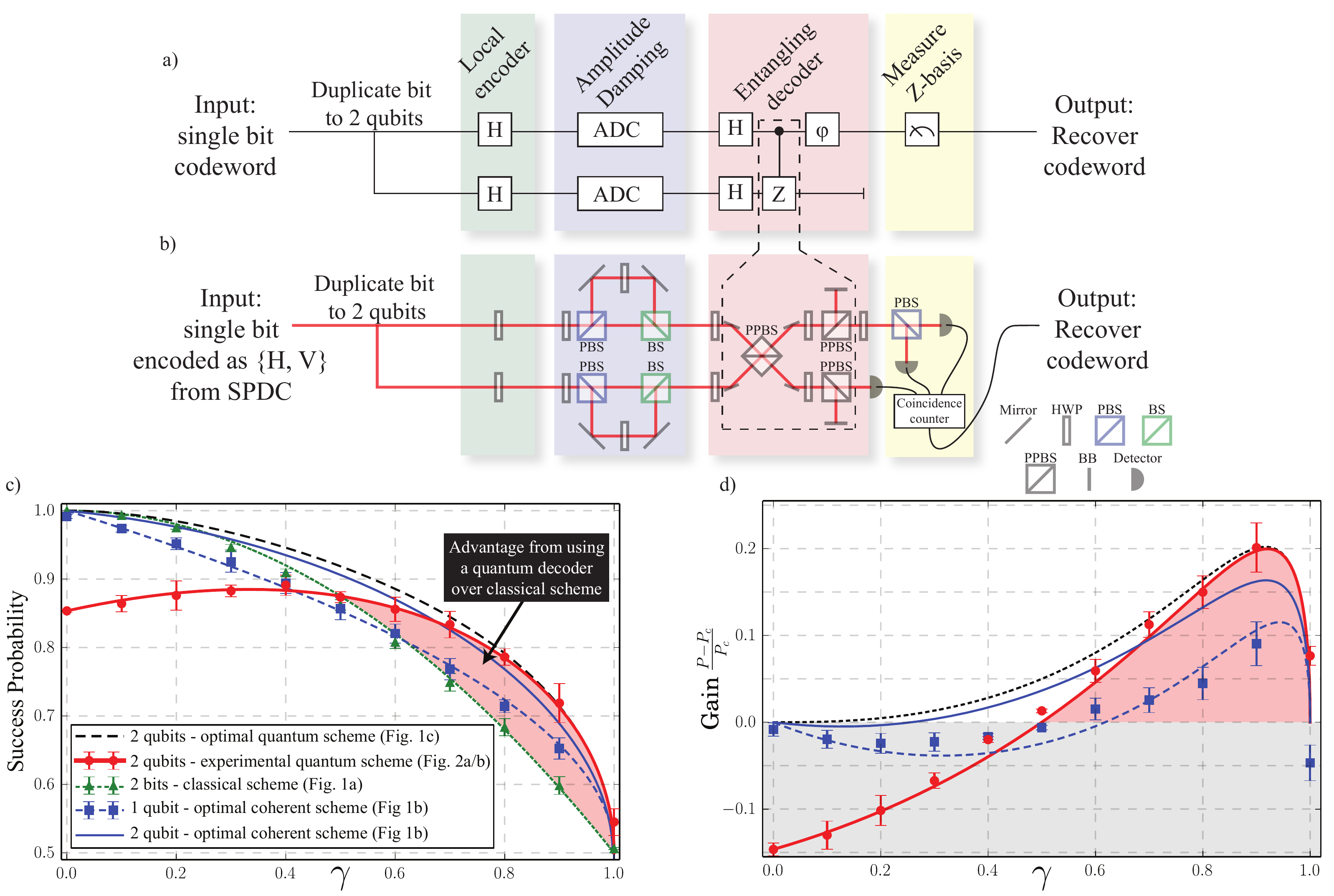}
\caption{
Experimentally implemented quantum scheme for transmitting one bit over an amplitude damping channel encoded in the state of two qubits.
(a) The circuit schematic for the entangling decoder we implement. 
This circuit is approximately optimal for large $\gamma$.
(b) Our experimental implementation based on polarization photonic qubits.
The components used are as follows: HWP, half wave plate; PBS, polarizing beam splitter; BS, 50/50 beam splitter; BB, beam block; PPBS, partially polarizing beam splitter (100\% horizontal transmission, 66\% vertical transmission). The $\varphi$ rotation is a $\sqrt{H}$ gate.
(c) Experimental results for the classical, coherent, and quantum schemes. 
The points are experimentally measured results and lines are calculated from circuit simulations. 
We also plot the theoretical maximum as a black dashed line and the shaded area is where we measure a quantum enhancement in message recovery.
(d) The gain over the classical scheme calculated as $\tfrac{P-P_c}{P_c}$.
The points are calculated using experimental data from the classical, coherent, and quantum schemes.
The lines are calculated from the ideal curves.
The red area highlights the advantage of the quantum scheme.
The gray area is where the classical scheme achieves a higher success probability and therefore the gain is negative.
}
\label{fig:fig_2}
\end{figure*}

\begin{figure*}
\centering
\includegraphics[width=1.0\linewidth]{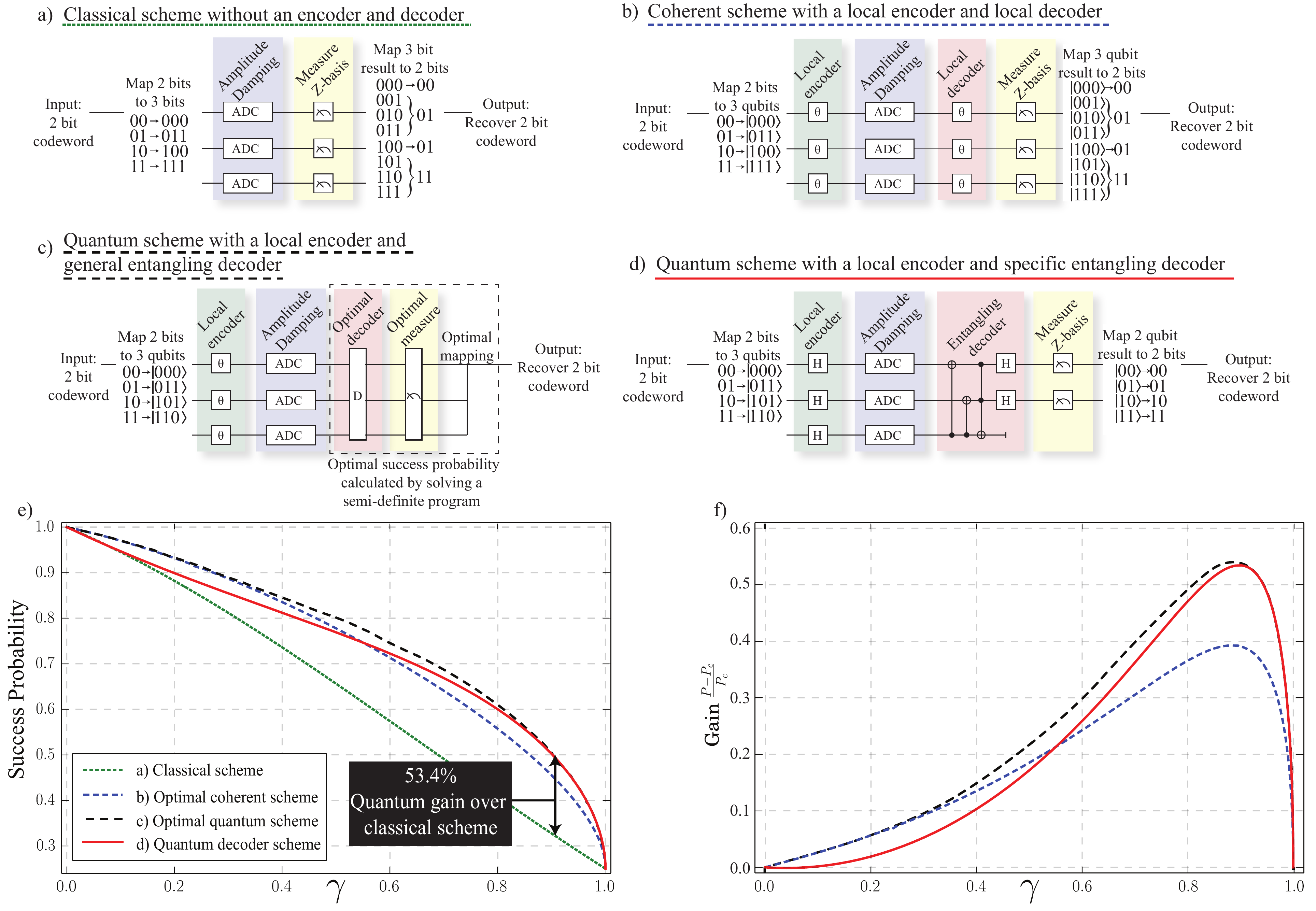}
\caption{
Numerical simulations for transmitting two bits over an amplitude damping channel encoded in the state of three qubits.
(a) The optimal classical scheme for transmitting a two bit message with three channel uses. 
(b) The optimal coherent scheme, where local encoding and decoding is used. 
(c) The maximum success probability with a quantum decoder. As there are more than two code words, the trace distance no longer directly calculates the optimal success probability. Therefore, a semi-definite program is solved to find the success probability for the optimal quantum decoder across all code words.
(d) A designed quantum decoder that approximates the optimal decoder for large $\gamma$.
(e) The success probability of each scheme. The entangling decoder surpasses the coherent scheme for $\gamma>0.55$ and achieves the theoretical maximum for large $\gamma$.
(f) The gain over the classical scheme for the coherent, analytic maximum and our quantum decoder schemes.
}
\label{fig:fig_3}
\end{figure*}

\section{Experimentally enhancing information transmission with a quantum decoder}

We use numerical optimization to find a gate sequence for transmitting each bit as two duplicate qubits which is near optimal for large $\gamma$ as this is the regime of greatest quantum gain (see Appendix 1 for further details).
The circuit designed is shown in Fig. \ref{fig:fig_2}(a) and consists of encoding both qubits with Hadamard gates before the ADC and decoding with a controlled phase gate along with local rotations. 
It is important to note that this decoder only requires the first qubit to be measured to recover the encoded information; however, in general the optimal decoder will require all qubits to be measured.
We experimentally construct this circuit for polarization photonic qubits as shown in Fig. \ref{fig:fig_2}b, where we prepare horizontally polarized photons from a type-1 spontaneous parametric down-conversion source and we follow the convention that horizontal (vertical) polarization corresponding to the state $\ket{0}$ ($\ket{1}$) (see Appendix 2 for details of the source).
We apply Hadamard encoding with half-wave plates (HWPs) and the ADC using an unbalanced interferometer, where a controllable portion of the photon wave packet is delayed beyond its coherence length and rotated to horizontal polarization.
This operation implements an ADC on the photonic polarization state (see Ref. \cite{fisher_optimal_2012, *qing_linear_2007, *lee_experimental_2011} for other optical implementations).
The entangling decoder is constructed with HWP rotations and a polarization controlled phase gate which works in post-selection with $\sfrac{1}{9}$ success probability, however, is heralded as successful when both photons are detected at the output \cite{langford_demonstration_2005, *kiesel_linear_2005, *okamoto_demonstration_2005}.
We characterize the ADC by performing polarization tomography after the channel on a range of input states and measure the average fidelity to the expected states of $96.6\pm0.2\%$. 
Repeating this process for the whole circuit including ADC and entangling decoder, we measure an average fidelity of $93.4\pm1.5\%$. 
The reduced fidelity of the ADC can be attributed to imprecision of each wave plate as well as imperfect mode overlap of the main and delayed portions of the wave packet.
For the entangling decoder, the main impact is from imperfect reflectivites of each partially polarizing beam splitter (PPBS) and the overlap of the two photons on the first PPBS leading to reduced quantum interference.
A $Z$-basis projection is performed with a polarizing beam splitter (PBS) on one photon before detection with silicon avalanche photodiodes and time correlated with a counting card.
The second photon is detected to herald the successful operation of the controlled-phase gate (see Appendix 3 for full details of the experimental setup).

Figure \ref{fig:fig_2}(c) presents the experimental results.
Error bars are calculated with a 95\% confidence by repeating each $\gamma$ value five times and each measurement uses around 2000 samples.
The lines are ideal curves from circuit simulations.
The green points show the experimental classical scheme success probability when simply duplicating the classical data.
The blue points are the experimental results for encoding the data in the $\{\ket{+},\ket{-}\}$ basis, which is optimal for a single-qubit coherent state.
Duplicating the message with this scheme does not increase the success probability as, in this basis, the damping is symmetric. 
However, with two uses of the channel and encoding with an angle of $\sim0.177\pi$ the maximum success probability is achievable, plotting as a solid blue line.
The red points in Fig. \ref{fig:fig_2}(c) show the results for the implemented quantum decoder scheme.

Figure \ref{fig:fig_2}(d) presents the percentage gain of each scheme over the classical scheme.
The points are calculated considering the experimental implementation of the classical scheme and the lines are the ideal curves.
Including error bars, our decoder surpasses the classical scheme for $\gamma>0.6$ and achieves a maximum relative increase of $20.1\pm1.2\%$ at $\gamma=0.9$.
Our quantum decoder achieves an advantage of $10.2\pm0.6\%$ at $\gamma=0.9$ over the optimal single channel use coherent scheme.
Compared to the optimal coherent scheme with two uses of the channel, our experimental quantum decoder still achieves up to $3.9\pm1.3\%$ higher success probability at $\gamma=0.9$.
In Figs.~\ref{fig:fig_2}(c) and \ref{fig:fig_2}(d) we have also plotted the two-qubit maximum success probability, calculated via the trace distance, as a black dashed line and our decoder matches the analytic maximum in the large-$\gamma$ regime.
A result of approximating the optimal quantum decoder in the large-$\gamma$ regime is that our simplified decoder is suboptimal at small $\gamma$ and drops below the classical scheme.

We have experimentally demonstrated that entangling two copies of a data qubit after an ADC can enhance the probability of recovering the encoded classical information.
Using adaptive techniques, where the result of projecting the first qubit influences the projective measurement of the second qubit can further enhance the success probability with the coherent scheme. 
However, this assumes additional control on the measurement which we leave for future work.
Our decoder only requires a single entangling gate beyond the coherent scheme, and a total of six gates beyond the classical scheme.
This resource overhead is modest given that we can achieve greater than $20\%$ gain over the classical scheme for a certain $\gamma$ range.
We have investigated using an entangling encode before the ADC and found this achieves a slightly lower success probability than an entangling decoder.
Additionally, if we include entangling gates in both the encoder and decoder, we achieve the same result as just the entangling decoder (see Appendix 4 for further details).

\section{Theoretical success probability for transmitting a two-bit message over an amplitude damping channel}

We next expand our methodology to enhance the success probability of sending a two-bit message $\{(00), (01), (10), (11)\}$ over an ADC with three uses of the channel.
We again consider four schemes: a classical scheme shown in Fig. \ref{fig:fig_3}(a), where the input and output mapping has been optimized, a coherent scheme shown in Fig. \ref{fig:fig_3}(b) with optimized mapping and local rotations, the optimal quantum scheme shown in Fig. \ref{fig:fig_3}(c), where a semidefinite program is solved to find the success probability of the optimal entangling decoder, measurement, and mapping, and finally a gate sequence with an entangling decoder that approximates the optimal scheme for large $\gamma$, shown in Fig. \ref{fig:fig_3}(d).
The success probabilities for all schemes are plotted in Fig. \ref{fig:fig_3}(e), showing that the quantum scheme achieves the analytic maximum for large $\gamma$, beats the classical scheme for $\gamma>0.079$ and the coherent scheme for $\gamma>0.55$.
Figure~\ref{fig:fig_3}(f) shows the gain of each scheme over the classical scheme.
The maximum gain of our quantum decoder over the classical scheme is 53.4\% at $\gamma=0.9$ and 10.5\%  gain over the coherent scheme at $\gamma=0.925$.
This large increase in success probability is the result of only a single additional channel use.
For these schemes, the mapping between the input message and transmitted code words is less trivial than the earlier one-bit case.
We consider all possible mapping schemes in our calculations; however, more efficient methods to find the optimal mapping would be beneficial.
Indeed for all encoding protocols that use redundancy, finding the optimal mapping between the data and transmitted code words becomes a challenge for large messages.

We have proposed and experimentally demonstrated a scheme for enhancing message recovery over an ADC by using a small amount of redundancy and an entangling decoder after the noisy channel.
We have demonstrated a two-qubit scheme to transmit a single classical bit over an ADC with greater than $20\%$ higher success probability than the optimal classical scheme.
Our protocol does not require the sender and receiver to share entanglement, or have access to additional noiseless channels, and only the receiver is required to have entangling capabilities.
For transmitting a two-bit message with three uses of the channel, our protocol demonstrated a greater than $50\%$ enhancement over the optimal classical scheme.
Important future investigations will extend our methodology to larger code words while using a single additional qubit to enhance the message recovery probability.

\section*{Acknowledgments}

We thank Tim Ralph and Nicolas Menicucci for helpful discussions. S.T.F. and M.T. acknowledge support from the Australian Research Council Centre of Excellence for Engineered Quantum Systems, Project No. CE110001013. S.T.F acknowledges an Australian Research Council Future Fellowship, Project No. FT130101744. M.T. acknowledges an Australian Research Council Discovery Early Career Researcher Award, Project No. DE160100821. A.P. acknowledges support from the Australian Research Council Centre of Excellent for Quantum Computation and Communication Technology (CQC$^2$T), Project No. CE170100012, an Australian Research Council Discovery Early Career Researcher Award, Project No. DE140101700 and an RMIT University Vice-Chancellor's Senior Research Fellowship.

\setcounter{equation}{0}
\makeatletter 
\renewcommand{\theequation}{A\@arabic\c@equation}
\makeatother

\section*{Appendix}

\subsection*{1: Theoretical Derivations} \label{supp:theory}

This appendix discusses our theoretical contribution. First, we give an upper bound on the success probability of any coding scheme that encodes and decodes for each quantum channel separately, and thus effectively embeds the quantum channel into the framework of classical information theory. Secondly, we discuss how one can beat this bound on the success probability by using a quantum decoding scheme jointly measuring two channel outputs.

\subsubsection*{Bound on the optimal success probability}

The following formula is from~\cite[Thm.~40]{polyanskiythesis10}, and corresponds to the sphere-packing bound~\cite[Eq.\ 5.8.19]{gallager68}. Consider a BSC with crossover probability $\delta$. Consider any $(n, M)$-code, where $M$ is the number of distinct messages we want to send and $n$ is the number of channel uses. Then the probability of decoding failure, $\eps$, must satisfy the following: 
\begin{align} \label{eq:M1}
  (1-\lambda) \beta_L + \lambda \beta_{L+1} \leq \frac{1}{M} , \quad \beta_{\ell} = \sum_{k=0}^{\ell} {n \choose k} 2^{-n} 
\end{align}
where the constants $\lambda$ and $L$ are determined by the relation
\begin{align} \label{eq:m2}
  1 - \eps &= (1-\lambda) \alpha_L + \lambda \alpha_{L+1}, \\ 
  \alpha_{\ell} &= \sum_{k=0}^{\ell} {n \choose k} (1-\delta)^{n-k} \delta^k \,.
\end{align}

Let us restate this for our case where $M$ is fixed, as is the case in our analysis here. In this case, Eq. \ref{eq:M1} determines $\lambda$ and $L$, i.e.,\ we want to find $\lambda$ and $L$ such that Eq. \ref{eq:M1} holds with equality. This is simple, since $\beta_L$ gradually increases with $L$, we simply need to find $L$ such that $\beta_L \leq \frac{1}{M} \leq \beta_{L+1}$.
Then, the (optimal) tradeoff between $\eps$ and $\delta$ is given by Eq. \ref{eq:m2} for the $\lambda$ and $L$ we have determined.

Here we want to transmit a 1 bit message over two uses of the channel. We can neatly arrange all the possible channel inputs, $\{(00), (01), (10), (11)\}$, on corners of a rectangle so that code words with distance $1$ are connected by an edge. We now choose two code words that lie mutually diagonal from each other, for example,
\begin{align}
  \{M_0, M_1\} = \{(00), (11)\} \,.
\end{align}
These are convenient because they have Hamming distance $2$ from each other. The channel will act as follows: it will flip each bit with probability $\delta$. For example, if we start at $(00)$ the probability of arriving at $(01)$ is $\delta (1-\delta)$ and the probability of arriving at $(11)$ is $\delta^2$. Our decoding procedure is very simple. When the channel output is $M_i$, then we will decode to $i$. If the channel output is not one of the code words, then we will decode to one of the two code words with Hamming distance $1$ of the channel output uniformly at random.

Clearly, this scheme always succeeds if no error occurs on the channel, and it succeeds with probability $\frac12$ if exactly one error occurs. The failure probability thus satisfies
\begin{align}
  \eps = 1 - (1-\delta)^2 - 2 \cdot \frac12 \delta (1-\delta) = \delta \,.
\end{align}

But is this optimal?
Going back to the calculation of the previous section, let us first note that $(\beta_0, \beta_1, \beta_2) = (\frac{1}{4}, \frac{3}{4}, 1)$ in this case. So, in order to satisfy Eq. \ref{eq:M1} for $M = 2$, we need to set $L = 0$ and $\lambda = \frac12$.
Hence we can evaluate 
\begin{align}
  \eps \geq 1 - (1-\delta)^2 - \delta (1-\delta) = \delta \,.
\end{align}
Hence we have shown that this code is optimal for $n = M = 2$.

\subsubsection*{Finding an efficient quantum coding scheme}

A coding scheme consists of two parts, an encoder preparing an input state depending on the message that is to be sent, and a decoder that attempts a state discrimination between the respective output states in order to decode the message.

\subsubsection*{Encoder: finding suitable input states}

In this work we consider only coding schemes using two independently prepared (product) input states. 
Schumacher and Westmoreland~\cite{schumacher01} determined the pair of input states that achieve capacity (for asymptotically many uses) of the amplitude damping channel. These are of the form
\begin{align}
  \ket{\pm}_{\alpha} = \sqrt{\alpha} \ket{0} \pm \sqrt{1-\alpha} \ket{1}
\end{align}
where $\alpha \in [\frac12, 1)$ and is usually very close to $\frac12$. These states remain optimal even when we consider a second order correction of the capacity formula~\cite{tomamichel_second-order_2015} that takes into account the finite size of the decoder. On the other hand, for a single use of the channel, we found in the previous section that $\alpha = \frac12$ is optimal. To see this, we just note that the trace distance between the outputs of the amplitude damping channel is maximized for the diagonal states, which we simply denote by $\ket{\pm}$ hereafter.

For our setup with two channel uses we find that the choice $\alpha = \frac12$ is sufficiently close to optimal. We thus fix our encoder to be the following map:
\begin{align}
 (0) &\mapsto \ket{\phi_0} = \ket{+} \otimes \ket{+} \\
 (1) &\mapsto \ket{\phi_1} = \ket{-} \otimes \ket{-} \,.
\end{align}
This means that if we want to transmit the message (0) we will prepare the state $\ket{\phi_0}$ and if we want to send the message (1) we will prepare the state $\ket{\phi_1}$.

The amplitude damping channel, $\cE_{\textrm{ad}}^{\gamma}$ , is then applied to these states. The output states are denoted
\begin{align}
  \rho_{\pm} = \cE_{\textrm{ad}}^{\gamma} \big(\ket{\pm}\!\bra{\pm} \big)
\end{align}
and the joint states corresponding to the messages (0) and (1) are
$\rho_{0} = \rho_{+} \otimes \rho_{+}$ and $\rho_{1} = \rho_{-} \otimes \rho_{-}$.

\subsubsection*{Decoder: approximately optimal decoder}

The task of the decoder is to distinguish between the states $\rho_0$ and $\rho_1$. In the most general framework of positive operator valued measures (POVMs), the decoder is determined by a positive semi-definite operator $0 \leq M_0 \leq 1$ acting on the two qubits. Let us say that $M_0$ indicates that the $\rho_0$ was detected [and thus (0) was sent]. We will also define $M_1 = 1 - M_0$ as its complement. 

The success probability of the decoder, if the two messages are chosen uniformly at random, is given by
\begin{align}
  p_{\rm succ}(M) 
  &= \frac12 \big( \tr( \rho_0 M_0 ) + \tr (\rho_1 M_1 ) \big) \\
  &= \frac12 + \frac12 \tr \big( M_0 (\rho_0 - \rho_1) \big) \label{eq:sdp} \\
  &\leq \frac12 + \frac12 \| \rho_0 - \rho_1 \|_{\rm tr} \,.
\end{align}
Here, we used the trace norm to bound the maximal success probability from above~\cite{helstrom76}.

The optimal decoding POVM to distinguish this set of states can be found quite easily by solving a semidefinite program. Namely, we need to find the maximum over $M_0$ of the expression in Eq. \ref{eq:sdp} subject to the constraint $0 \leq M_0 \leq 1$.
The optimal POVM elements do not have a simply analytical form, making them difficult to implement in the lab. Generally, the optimal quantum decoder also depends on the amplitude damping parameter $\gamma$.

However, we find that the optimal decoder can be approximated very well, at least for sufficiently large values of $\gamma$, by a simple decoding circuit. [See Fig. \ref{fig:fig_2}(a) in the main text.]
The simpler, approximately optimal decoder is determined by the following set of projectors:
\begin{align}
   P_{k} &= |v_{0,k}\rangle\!\langle v_{0,k}| + |v_{1,k}\rangle\!\langle v_{1,k}|
\end{align}
with $|v_{j,k} \rangle = U(k) |v_j \rangle$ for $j \in \{0, 1\}$, and
\begin{align}
   |v_0\rangle &= \frac{1}{\sqrt{2}} |00\rangle + \frac12 \left( |01\rangle + |10\rangle \right), \label{eq:dec}\\
   |v_1\rangle &= \frac{1}{\sqrt{2}} |11\rangle + \frac12 \left( |01\rangle - |10\rangle \right) \notag.
\end{align}

We have seen that this nearly optimal rank-1 POVM decoder simply needs to measure in the basis given in Eq. \ref{eq:dec}. This measurement acts on two qubits and is thus nontrivial to implement experimentally. Here we show that such a measurement can be decomposed into simple gates. Let us first define an operator $V$, specified via the following circuit:
\begin{align}
V =\begin{array}{c} \Qcircuit @C=1.5em @R=1em {
	& \gate{H} & \ctrl{1} & \gate{\frac{\pi}{8}} & \qw & \qw \\
	& \qw      & \targ    & \gate{H}       & \gate{\frac{\pi}{8}} & \qw \\
}\end{array}
\end{align}
Namely, if we take the top-most qubit to be the most significant one, we find
\begin{align}
  &V^{\dag} |00\rangle = |v_0\rangle, \quad V^{\dag} |11\rangle = Z \otimes Z|v_0\rangle \\
  &V^{\dag} |01\rangle = |v_1\rangle, \quad V^{\dag} |10\rangle = Z \otimes Z|v_1\rangle
\end{align}
Now, it is evident that if we feed the output of our damping channel in this circuit from the left-hand side and then measure the topmost qubit (and ignore the bottom-most qubit), this in fact exactly implements the decoding measurement in Eq. \ref{eq:dec}. So the nearly optimal decoder is simply the circuit displayed in Fig. \ref{fig:fig_2}(a) in the main text.

\subsection*{2: Spontaneous Parametric Down Conversion Photon Pair Source}

\begin{figure}[t]
\centering
\includegraphics[width=0.8\linewidth]{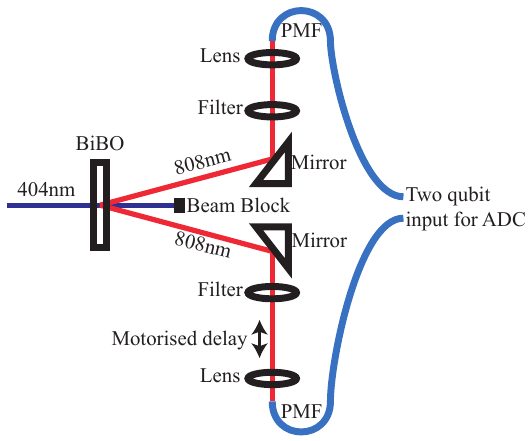}
\caption{Spontaneous parametric down conversion source of indistinguishable horizontally polarized photon pairs.}
\label{fig:supp_2}
\end{figure}

Horizontally polarized photon pairs at 807.5 nm are generated via type 1 spontaneous parametric down conversion (SPDC) in a 1-mm-thick BiBO crystal, pumped by an 80 mW, 403.75 nm cw diode laser \cite{burnham_observation_1970}. 3 nm FWHM filters are used on both photons to ensure near perfect indistinguishability in wavelength before coupling into polarization maintaining fiber. One fiber port is positioned on a motorized stage to enable the photon path lengths to be matched, which is crucial for the controlled phase gate which relies on photon bunching at the first partially polarizing beam splitter [see main text Fig. \ref{fig:fig_2}(b)]. Figure \ref{fig:supp_2} shows a schematic of the photon pair source.

\subsection*{3: polarization ADC and controlled phase gate setup}

We implement an ADC on the polarization state of the photon, where the $\ket{V}$ component damps to $\ket{H}$ (see Fig. 2 of the main text).
The SDPC source prepares pairs of $\ket{H}$ polarized photons, which are injected as $\ket{HH}$ to the ADC if we want to transmit a bit state $(0)$ and as $\ket{VV}$ if we want to transmit $(1)$.
Local Hadamard encoding is implemented for each photon with a half-wave plate (HWP).
The polarization ADC for each photon is constructed of HWPs, a polarizing beam splitter (PBS) and a 50/50 beam splitter (BS).
The first HWP controls the polarization of the photon before the PBS which spatially separates horizontal (transmission) and vertical (reflection). 
This forms the two arms of an unbalanced interferometer where the reflected component has a longer path length than the transmitted component before they are recombined at the BS.
The path length difference is greater than the coherence length of the photon, however, short enough that on detection this degree of freedom is traced out.
By suitably choosing the angles of the three HWPs in the ADC, it is possible to implement amplitude damping with any $\gamma$ value on the polarization photonic state.

The entangling decoder is based on HWPs and partially polarizing beam splitters (PPBSs) which transmit 100\% of horizontal polarization and 33\% of vertical.
The two photons are both incident at the first PPBS, causing partial Hong-Ou-Mandel interference. 
After the first PPBS, each photon has a Pauli-$X$ operation applied with a HWP before a second PPBS.
This PPBS is necessary to balance the probabilities of each component of the controlled-phase transfer matrix and the reflected components are removed with beam blocks.
In post-selection, when both photons are detected, this operation implements a polarization controlled-phase gate.
As with all linear-optical entangling gates, this operation is probabilistic and has a success rate of $\sfrac{1}{9}$; however, post-selection ensures successful operation for all recorded events \cite{ralph_linear_2002}.
This physically reduced the success rate of our decoding protocol; however, this is a manifestation of entangling gates in linear optics and is not a feature of the decoder.

A polarization $Z$-basis projection measurement is implemented with a PBS before silicon avalanche photodiodes. In our protocol the second qubit measurement is not required; however, we detect the second photon to herald the successful operation of the controlled-phase gate.

\subsection*{4: Entangling encoders vs. entangling decoders}

\begin{figure}[t]
\centering
\includegraphics[width=1.0\linewidth]{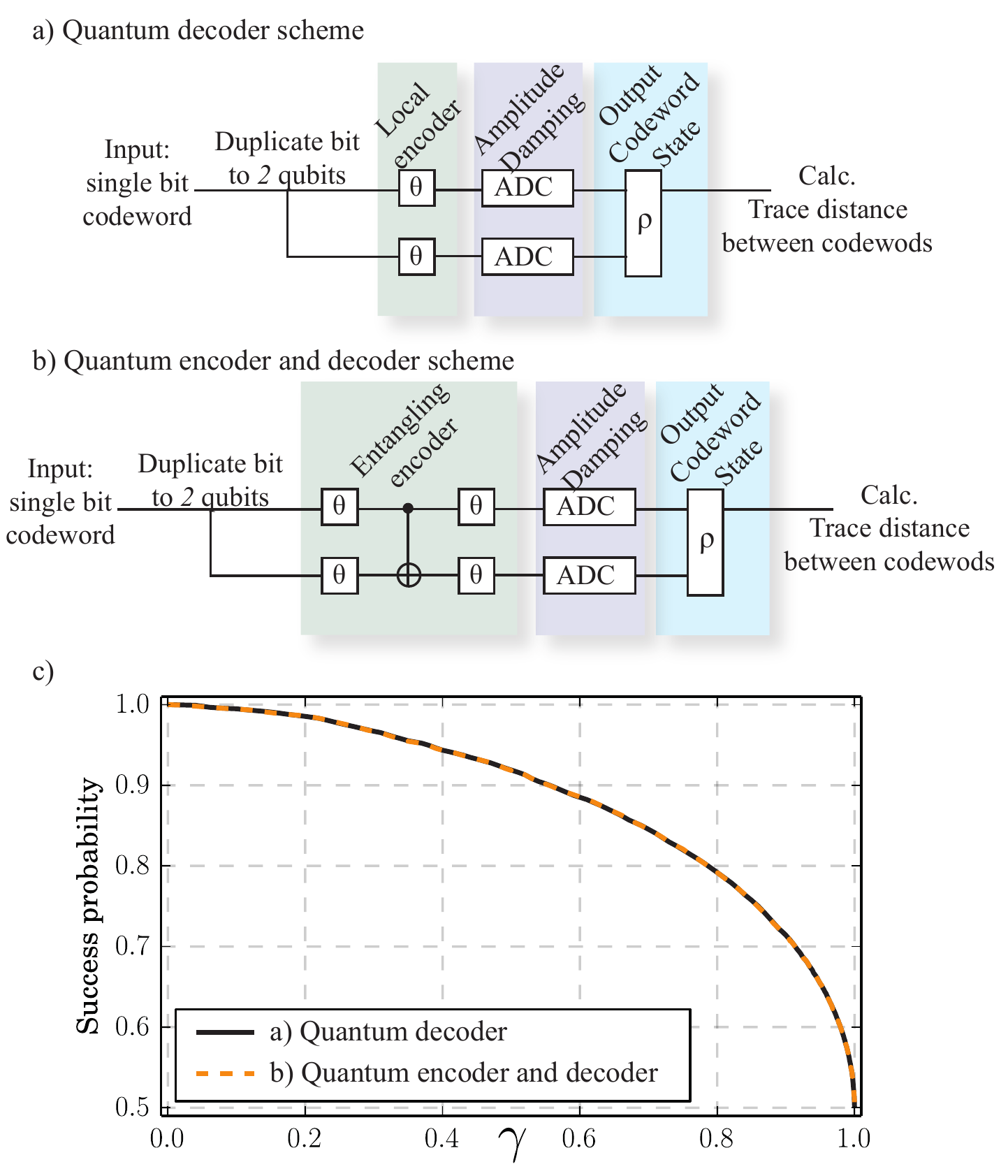}
\caption{(a) Quantum scheme including local encoder and entangling decoder for transmitting a one bit message over two qubits.
(b) A scheme including an entangling encoder and decoder.
(c) The success probability for each of these schemes, demonstrating that the entangling encoder adds no benefit to the success probability.}
\label{fig:supp_1}
\end{figure}

We have performed initial simulations on the benefit of including an entangling encoder as well as a decoder. For this investigation we focus on the case of sending one bit over the ADC with two qubits.
We use the trace distance and optimized local encoding as per the scheme shown in Fig. \ref{fig:supp_1}(a) which is equivalent to Fig. \ref{fig:fig_1}(c) of the main text.
We also include an entangling encoder, which we form with four local rotations surrounding a controlled-NOT (CNOT) gate, as shown in Fig. \ref{fig:supp_1}(b).
In Fig. \ref{fig:supp_1}(c) we plot the success probability for each of these schemes, where the local rotations are numerically optimized for each $\gamma$ value.
We find that the success probability is identical for both schemes, demonstrating that this entangling encoder adds no benefit to the communication scheme. We leave full investigation of this scheme for future work and in the main text focus only on the entangling decoder.

\begin{figure}[t]
\centering
\includegraphics[width=1.0\linewidth]{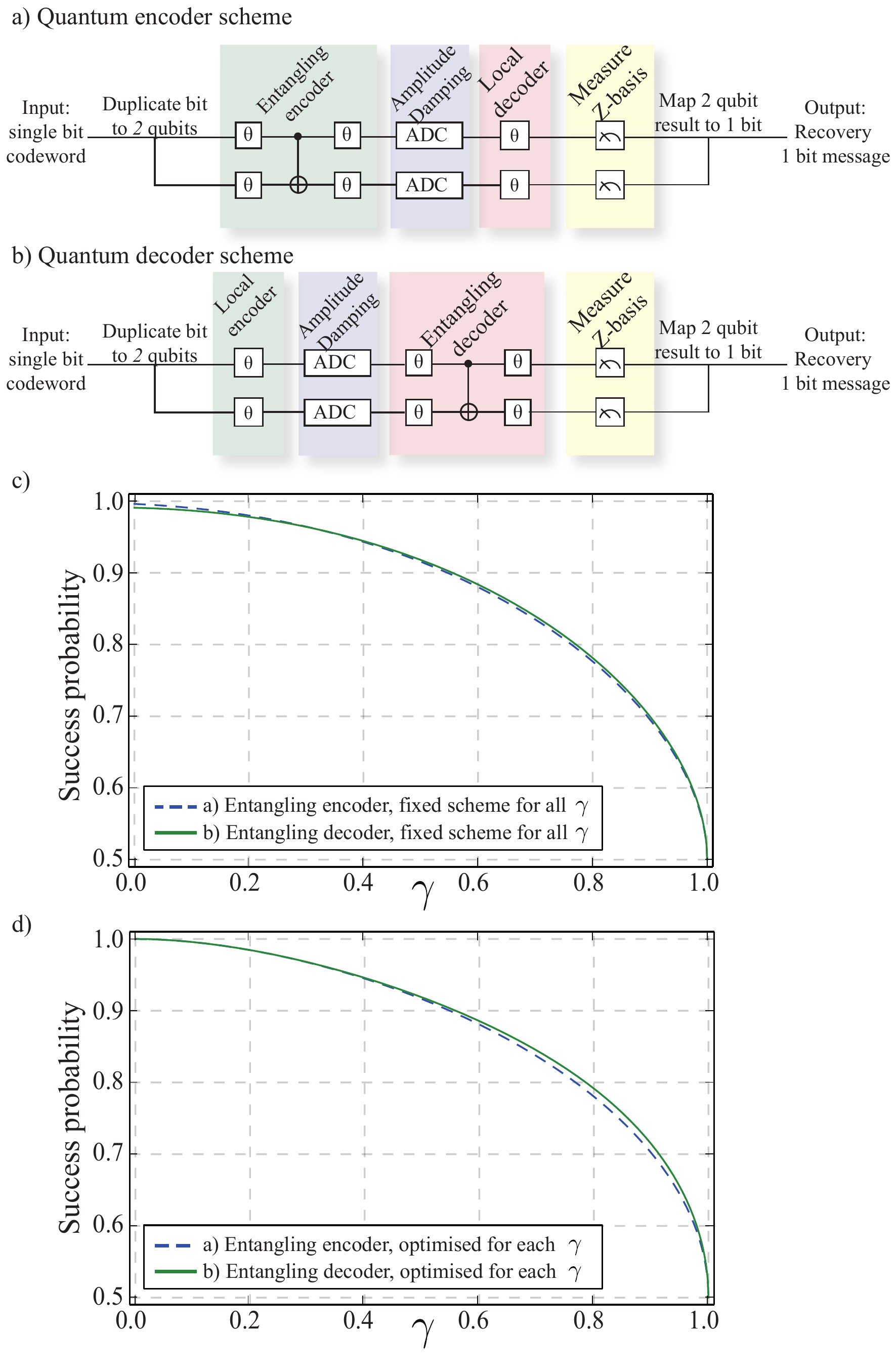}
\caption{(a) Quantum encoding scheme with a local decoder.
(b) A local encoding scheme with a quantum decoder.
(c) The success probability for each of these schemes where we optimize for all $\gamma$ values simultaneously, to maximise the average success probability.
(d) The success probability where we have optimized for each $\gamma$ value independently.}
\label{fig:supp_3}
\end{figure}

We next numerically compare having one entangling gate at either the encoder or decoder. 
For the entangling encoder, we consider the scheme shown in Fig. \ref{fig:supp_3}(a), where the encoder is a CNOT gate surrounded by four local rotations.
The decoder is a local rotation of each qubit before a $Z$-basis measurement and a mapping where if either qubit is measured as $\ket{1}$, then the original bit is decoded as (1).
The entangling decoder scheme is shown in Fig. \ref{fig:supp_3}(b), where the encoder is now local and decoder is a CNOT and local rotations.
We first numerically optimize the six local rotations in each scheme to find the maximum average success probability across the full range of $\gamma$.
The success probability of the optimal fixed schemes are shown in Fig. \ref{fig:supp_3}(c).
There is clearly little difference between the two schemes; however, the decoder achieves a slightly higher success probability for $\gamma\gtrsim0.3$.
For the optimization where we consider each $\gamma$ value individually shown in Fig. \ref{fig:supp_3}(d), the entangling decoder achieves higher than the encoder for larger $\gamma$ values.
The difference is still small, however, larger than the case for fixed schemes.

\end{document}